\documentclass[aps,pra,floatfix,showpacs,preprint,superscriptaddress]{revtex4-2}
\usepackage{bm}
\usepackage{mathrsfs}
\usepackage{amsmath}
\usepackage{amssymb}
\usepackage{revsymb}
\usepackage{accents}
\usepackage{braket}

\usepackage{color}
\usepackage[force]{feynmp-auto}
\usepackage{hyperref}
\usepackage{cleveref}

\usepackage{graphicx}

\begin{document}
\title{On the electron mass shift in an intense plane wave}
\author{A. \surname{Di Piazza}}
\email{dipiazza@mpi-hd.mpg.de}
\author{T. \surname{P\u{a}tuleanu}}
\affiliation{Max Planck Institute for Nuclear Physics, Saupfercheckweg 1, D-69117 Heidelberg, Germany}
\date{\today}

\begin{abstract}
Upcoming experiments on the interaction of electrons with intense laser fields are envisaged to become more and more accurate, which calls for theoretical computations of rates and probabilities with correspondingly higher precision. In strong-field QED this requires the knowledge of radiative corrections to be added to leading-order results. Here, we first derive the mass operator in momentum space of an off-shell electron in the presence of an arbitrary plane wave. By taking the average of the mass operator in momentum space over an on-shell electron state, we obtain a new representation, equivalent to but more compact than the known one computed in [Sov. Phys. JETP \textbf{42}, 400 (1975)]. Moreover, we use the obtained mass operator to determine the electron mass shift in an arbitrary plane wave, which generalizes the already known expression in a constant crossed field. The spin-dependent part of the electron mass shift can be related to the anomalous magnetic moment of the electron in the plane wave. We show that within the locally constant field approximation it is possible to conveniently define a local expression of the electron anomalous magnetic moment, which reduces to the known expression in a constant crossed field. Beyond the locally constant field approximation, however, the interaction between the electron and the plane wave is non-local such that it is not possible to conveniently introduce an electron anomalous magnetic moment.

  
\end{abstract}
 
\maketitle

\section{Introduction}
Among the most stringent experimental tests on QED the measurement of the anomalous magnetic moment of either a free \cite{Hanneke_2008,Hanneke_2011} or a bound \cite{Sturm_2011} electron plays certainly a prominent role. Also from an historical point of view the anomalous magnetic moment of the electron has a distinguished position with Schwinger's computation of the leading-order contribution representing the first successful application of covariant renormalization theory \cite{Schwinger_1948}.

In experimental conditions like those described in Ref. \cite{Sturm_2011} electrons bound in highly-charged ions experience electric fields of strengths of the order of the QED scale, the so-called critical electric field of QED $E_{cr}=m^2c^3/\hbar|e|\approx 1.3\times 10^{16}\;\text{V/cm}$, with $e<0$ and $m$ being the electron charge and mass, respectively \cite{Mitter_1975,Ritus_1985,Ehlotzky_2009,Reiss_2009,Di_Piazza_2012,Dunne_2014}. It is desirable, however, to test the theory also in the presence of fields with a different structure and with different properties. 

High-power lasers offer an alternative tool to test QED at the critical field scale. In fact, although the electric fields provided by existing and forthcoming facilities are well below the critical fields \cite{Yoon_2021,APOLLON_10P,ELI,CoReLS,Bromage_2019,XCELS}, observable quantities depend on the laser field amplitude $F_0^{\mu\nu}=(\bm{E}_0,\bm{B}_0)$ through the so-called quantum nonlinearity parameter $\chi_0=\sqrt{|(F_0^{\mu\nu}p_{\nu})^2|}/mE_{cr}$, where $p^{\mu}=(\varepsilon,\bm{p})$ is the four-momentum of the electron (or positron), with $\varepsilon=\sqrt{m^2+\bm{p}^2}$ (the metric tensor $\eta^{\mu\nu}=\text{diag}(+1,-1,-1,-1)$ and units with $\epsilon_0=\hbar=c=1$ are used) \cite{Mitter_1975,Ritus_1985,Ehlotzky_2009,Reiss_2009,Di_Piazza_2012,Dunne_2014}. Indeed, the parameter $\chi_0$ numerically corresponds to the laser electric field amplitude in the rest frame of the electron/positron in units of $E_{cr}$, which is boosted as compared to the value in the laboratory frame by a factor of the order of the electron relativistic Lorentz factor.

The first experiments on strong-field QED in the presence of intense laser radiation were carried out at SLAC in the late nineties \cite{Bula_1996,Burke_1997,Bamber_1999} and more recently two experiments have been performed \cite{Cole_2018,Poder_2018} (see also Refs. \cite{Wistisen_2018,Wistisen_2019} for similar experiments in a crystalline field). Moreover, devoted experimental campaigns are already planned at DESY \cite{Abramowicz_2019} and at SLAC \cite{Meuren_2020} to further test QED at background fields effectively of the order of $E_{cr}$ and beyond. The main differences between the first experiments at SLAC and the recent ones are that: 1) at SLAC an electron beam from a traditional accelerator was employed, whereas recently electron bunches produced via laser wakefield acceleration were used, and 2) at SLAC the so-called classical nonlinearity parameter $\xi_0=|e|E_0/m\omega_0$, where $\omega_0$ is the central angular frequency of the laser field, was smaller than unity, whereas in recent experiments it well exceeds unity. The second difference implies that nonlinear effects in the laser amplitude are much more pronounced in the late experiments because even an electron at rest would be accelerated to relativistic energies in a fraction of the laser period \cite{Mitter_1975,Ritus_1985,Ehlotzky_2009,Reiss_2009,Di_Piazza_2012,Dunne_2014}. However, electron beams produced by traditional accelerators are presently still more stable and have a higher quality in terms of monochromaticity and emittance than laser-produced beams, and allow for more precise measurements. This is one of the aims of the mentioned future campaigns planned at DESY \cite{Abramowicz_2019} and at SLAC \cite{Meuren_2020}, which will again use high-quality electron beams and allow for precise measurements also in the nonlinear regime where $\xi_0>1$.

In view of the increasing accuracy of experiments testing QED in the strong-field regime with high-intensity lasers, it is appropriate from a theoretical point of view to start investigating radiative corrections of the basic processes, which have been studied in detail over the last years, namely, nonlinear Compton scattering \cite{Goldman_1964,Nikishov_1964,Ritus_1985,Baier_b_1998,Ivanov_2004,Boca_2009,
Harvey_2009,Mackenroth_2010,Boca_2011,Mackenroth_2011,Seipt_2011,Seipt_2011b,
Dinu_2012,Krajewska_2012,Dinu_2013,Seipt_2013,
Krajewska_2014,Wistisen_2014,Harvey_2015,Seipt_2016,Seipt_2016b,Angioi_2016,
Harvey_2016b,Angioi_2018,Di_Piazza_2018,Alexandrov_2019,Di_Piazza_2019,Ilderton_2019_b} and nonlinear Breit-Wheeler pair production \cite{Reiss_1962,Nikishov_1964,Narozhny_2000,Roshchupkin_2001,Reiss_2009,
Heinzl_2010b,Mueller_2011b,Titov_2012,Nousch_2012,Krajewska_2013b,Jansen_2013,
Augustin_2014,Meuren_2016,Di_Piazza_2019,King_2020}. Due to technical difficulties arising in order to take into account exactly the background laser field \cite{Furry_1951,Fradkin_b_1991,Landau_b_4_1982}, in the mentioned works the laser field has been described by a plane wave, an approximation valid if the laser field is not too tightly focused. There are theoretical schemes, which allow to investigate strong-field QED processes taking into account the complex spacetime structure of the background laser field like the so-called locally constant field approximation (LCFA) \cite{Ritus_1985,Baier_b_1998}, Baier's semiclassical operator approach \cite{Baier_1967,Baier_1968,Baier_1969,Baier_b_1998}, and a systematic approach based on the  Wentzel-Kramers-Brillouin (WKB) approximation \cite{Di_Piazza_2014,Di_Piazza_2015,Di_Piazza_2016,Di_Piazza_2017_b,Di_Piazza_2021_c}. 

\begin{figure}
\begin{center}
\includegraphics[width=0.6\columnwidth]{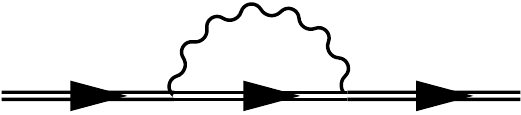}
\caption{The one-loop mass operator in an intense plane wave. The double lines represent exact electron states and propagator in the plane wave (Volkov states and propagator, respectively) \cite{Landau_b_4_1982}.}
\label{FD_MO}
\end{center}
\end{figure}

\begin{figure}
\begin{center}
\includegraphics[width=0.6\columnwidth]{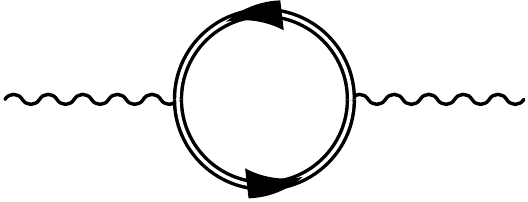}
\caption{The one-loop polarization operator in an intense plane wave. The double lines represent exact electron propagators in the plane wave (Volkov propagators) \cite{Landau_b_4_1982}.}
\label{FD_PO}
\end{center}
\end{figure}

\begin{figure}
\begin{center}
\includegraphics[width=0.4\columnwidth]{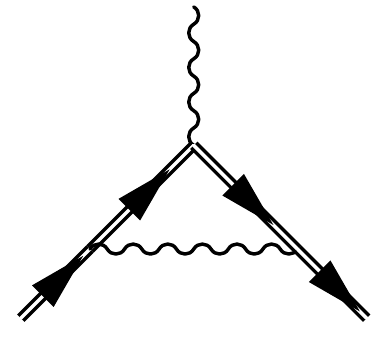}
\caption{The one-loop vertex correction in an intense plane wave. The double lines represent exact electron states and propagator in the plane wave (Volkov states and propagator, respectively) \cite{Landau_b_4_1982}.}
\label{FD_VC}
\end{center}
\end{figure}

The computation of radiative corrections, due to intrinsic technical difficulties, has been only carried out so far within the plane-wave approximation, meaning that the electron states have been employed, which are solutions of the Dirac equation in the presence of a plane-wave field (Volkov states \cite{Volkov_1935}), as well as the corresponding electron propagator. The one-loop mass operator (see Fig. \ref{FD_MO}) and the one-loop polarization operator (see Fig. \ref{FD_PO}) in the presence of an arbitrary plane wave have been computed in Ref. \cite{Baier_1976_a} and in Refs. \cite{Becker_1975,Baier_1976_b,Meuren_2013}, respectively.
Finally, the one-loop vertex correction (see Fig. \ref{FD_VC}) was recently computed in Ref. \cite{Di_Piazza_2020_b}.

A more systematic investigation has been carried out in the case of a constant crossed field, especially in relation to the so-called Ritus-Narozhny conjecture, stating that at $\chi_0\sim 1/\alpha^{3/2}\gg 1$, where $\alpha=e^2/4\pi$ is the fine-structure constant, and $\xi_0^3\gg \chi_0$ the perturbative approach to strong-field QED breaks down as the coupling constant in this regime is not $\alpha$ but rather $\alpha\chi_0^{2/3}$ \cite{Ritus_1970,Narozhny_1979,Narozhny_1980,Morozov_1981,Akhmedov_1983,Akhmedov_2011,Fedotov_2017,
Podszus_2019,Ilderton_2019,Mironov_2020}.

In the present paper, we compute the mass operator in momentum space of an off-shell electron in the presence of an arbitrary plane wave. We point out that in Ref. \cite{Baier_1976_a} only an operator form of the mass operator is presented in the general, off-shell case. Then, by averaging over an on-shell Volkov state, the electron mass correction was also obtained in Ref. \cite{Baier_1976_a}. The mass operator in momentum space is defined here in such a way that the same electron mass correction is computed by averaging the obtained mass operator in momentum space over a free on-shell electron state, and the resulting expression turns out to be equivalent but more compact than that found in Ref. \cite{Baier_1976_a}. Then, we use the spin-dependent part of the mass shift in the case of a linearly polarized plane wave to study the anomalous magnetic moment of the electron. We show that within the LCFA it is possible to introduce a local expression of the anomalous magnetic moment of the electron, which reduces to the known one in a constant crossed field, already computed in Refs. \cite{Ritus_1970,Baier_1971}. In the case of an arbitrary plane wave, however, the electron mass shift features a non-local dependence on the plane-wave field, which prevents a convenient description of the spin-dependent part in terms of an electron anomalous magnetic moment.

\section{The one-loop mass operator in a general plane wave}

In this section we compute the one-loop mass operator of an electron in a general plane wave. Before we introduce some basic definitions.

\subsection{Basic definitions}
For the sake of definiteness, we consider a plane wave propagating along the direction $\bm{n}$ and thus being described by the four-vector potential $A^{\mu}(\phi)$, where $\phi=t-\bm{n}\cdot \bm{x}$ is the so-called light-cone time. It is convenient to introduce a basis of the four-vector space starting from the two quantities $n^{\mu}=(1,\bm{n})$ and $\tilde{n}^{\mu}=(1,-\bm{n})/2$ (note that $\phi=(nx)$). By introducing two additional four-vectors $a_j^{\mu}=(0,\bm{a}_j)$, with $j=1,2$, such that $(na_j)=-\bm{n}\cdot\bm{a}_j=0$ and $(a_ja_{j'})=-\bm{a}_j\cdot\bm{a}_{j'}=-\delta_{jj'}$, with $j,j'=1,2$, it is clear that the completeness relation $\eta^{\mu\nu}=n^{\mu}\tilde{n}^{\nu}+\tilde{n}^{\mu}n^{\nu}-a_1^{\mu}a_1^{\nu}-a_2^{\mu}a_2^{\nu}$ holds (note that $(n\tilde{n})=1$ and $(\tilde{n}a_j)=0$). Also, the four-vector potential $A^{\mu}(\phi)$ can be chosen to fulfill the Lorenz condition $\partial_{\mu}A^{\mu}=0$, with the additional constraint $A^0(\phi)=0$, and it is also assumed to fulfill the boundary conditions $\lim_{\phi\to\pm\infty}A^{\mu}(\phi)=0$. This allows one to represent $A^{\mu}(\phi)$ in the form $A^{\mu}(\phi)=(0,\bm{A}(\phi))$, where $\bm{n}\cdot\bm{A}(\phi)=0$. Thus, the vector $\bm{A}(\phi)$ can be written as $\bm{A}(\phi)=\psi_1(\phi)\bm{a}_1+\psi_2(\phi)\bm{a}_2$, where the two functions $\psi_j(\phi)$ are arbitrary, provided that they vanish for $\phi\to\pm\infty$ and they feature obvious differential properties. Below, we will refer to the transverse ($\perp$) plane as the plane spanned by the two perpendicular unit vectors $\bm{a}_j$. Thus, together with the light-cone time $\phi=t-\bm{n}\cdot \bm{x}$, we also introduce the remaining three light-cone coordinates $\tau=(\tilde{n}x)=(t+\bm{n}\cdot \bm{x})/2$, and $\bm{x}_{\perp}=(x_{\perp,1},x_{\perp,2})=-((xa_1),(xa_2))=(\bm{x}\cdot\bm{a}_1,\bm{x}\cdot\bm{a}_2)$. Analogously, the light-cone coordinates of an arbitrary four-vector $v^{\mu}=(v_0,\bm{v})$ will be indicated as $v_-=(nv)=v_0-\bm{n}\cdot \bm{v}$, $v_+=(\tilde{n}v)=(v_0+\bm{n}\cdot \bm{v})/2$, and $\bm{v}_{\perp}=(v_{\perp,1},v_{\perp,2})=-((va_1),(va_2))=(\bm{v}\cdot\bm{a}_1,\bm{v}\cdot\bm{a}_2)$.

The field tensor $F^{\mu\nu}(\phi)=\partial^{\mu}A^{\nu}(\phi)-\partial^{\nu}A^{\mu}(\phi)$ of the plane wave is given by $F^{\mu\nu}(\phi)=n^{\mu}A^{\prime\,\nu}(\phi)-n^{\nu}A^{\prime\,\mu}(\phi)$, whereas the dual field tensor by $\tilde{F}^{\mu\nu}(\phi)=(1/2)\varepsilon^{\mu\nu\lambda\rho}F_{\lambda\rho}(\phi)$, with $\varepsilon^{\mu\nu\lambda\rho}$ being the four-dimensional Levi-Civita symbol, with $\varepsilon^{0123}=+1$. Here and below, the prime in a function of $\phi$ indicates the derivative with respect to $\phi$.

Since the four-vector potential $A^{\mu}(\phi)$ of the plane wave very often appears multiplied by the electron charge $e$, it is convenient to introduce the notation $\mathcal{A}^{\mu}(\phi)=eA^{\mu}(\phi)$ and, correspondingly, $\mathcal{F}^{\mu\nu}(\phi)=eF^{\mu\nu}(\phi)$ and $\tilde{\mathcal{F}}^{\mu\nu}(\phi)=e\tilde{F}^{\mu\nu}(\phi)$.

The Dirac equation in the presence of a plane wave can be solved exactly and, as we have already mentioned, the corresponding states are known as Volkov states \cite{Volkov_1935,Landau_b_4_1982} (see also Ref. \cite{Di_Piazza_2021_c} for a new, alternative form of the Volkov states). Below, the four-vector $p^{\mu}=(\varepsilon,\bm{p})$ indicates an on-shell electron four-momentum, i.e., $\varepsilon=\sqrt{m^2+\bm{p}^2}$, and $\sigma$ indicates the spin quantum number. The latter refers for the moment to an arbitrary quantization axis and a specific choice will be made in Sec. \ref{EMS}. The positive-energy Volkov state $U_{\sigma}(p,x)$ can be written in the form $U_{\sigma}(p,x)=E(p,x)u_{\sigma}(p)$, where
\begin{equation}
\label{E_p}
E(p,x)=\bigg[1+\frac{\hat{n}\hat{\mathcal{A}}(\phi)}{2p_-}\bigg]e^{i\left\{-(px)-\int_0^{\phi}d\varphi\left[\frac{(p\mathcal{A}(\varphi))}{p_-}-\frac{\mathcal{A}^2(\varphi)}{2p_-}\right]\right\}}
\end{equation}
and where $u_{\sigma}(p)$ is the free, positive-energy spinor normalized as $u^{\dag}_{\sigma}(p)u_{\sigma'}(p)=2\varepsilon \delta_{\sigma \sigma'}$ \cite{Landau_b_4_1982}. In Eq. (\ref{E_p}) we have introduced the notation $\hat{v}=\gamma^{\mu}v_{\mu}$, where $\gamma^{\mu}$ are the Dirac matrices and $v^{\mu}$ is a generic four-vector. Note that the expression in Eq. (\ref{E_p}) can also formally be used for the matrix $E(l,x)$, where $l^{\mu}=(l^0,\bm{l})$ is a generic off-shell four-momentum, and this matrix fulfills the identities \cite{Ritus_1985,Di_Piazza_2018_d}:
\begin{align}
\label{Compl_x}
\int d^4x \bar{E}(l,x)E(l',x)&=(2\pi)^4\delta^4(l-l'),\\
\label{Compl_l}
\int \frac{d^4l}{(2\pi)^4} E(l,x)\bar{E}(l,y)&=\delta^4(x-y),\\
\label{E_p_Comm}
\gamma^{\mu}[i\partial_{\mu}-\mathcal{A}_{\mu}(\phi)]E(l,x)&=E(l,x)\hat{l},
\end{align}
where $l^{\prime\,\mu}=(l^{\prime\,0},\bm{l}')$ is another off-shell four-momentum and where, for a generic matrix $O$ in the Dirac space, we have introduced the notation $\bar{O}=\gamma^0O^{\dag}\gamma^0$. 

By means of the matrices $E(l,x)$ one can also define the Volkov propagator $G(x,y)$ as  \cite{Ritus_1985,Di_Piazza_2018_d}
\begin{equation}
G(x,y)=\int \frac{d^4l}{(2\pi)^4}E(l,x)\frac{\hat{l}+m}{l^2-m^2+i0}\bar{E}(l,y)
\end{equation}
but below we will rather use the operator representation of the Volkov propagator, where $G(x,y)=\braket{x|G|y}$, with $\ket{x}$ and $\ket{y}$ being the eigenstates of the four-position operator $X^{\mu}$ with eigenvalue $x^{\mu}$ and $y^{\mu}$, respectively, and  with \cite{Baier_1976_a,Baier_1976_b,Di_Piazza_2007}
\begin{equation}
\label{G}
\begin{split}
G=&\frac{1}{\hat{\Pi}(\Phi)-m+i0}=\frac{1}{\hat{\Pi}^2(\Phi)-m^2+i0}[\hat{\Pi}(\Phi)+m]\\
&=(-i)\int_0^{\infty}ds\, e^{-im^2s}\Big\{1+\frac{1}{2P_{\tau}}\hat{n}[\hat{\mathcal{A}}(\Phi+2sP_{\tau})-\hat{\mathcal{A}}(\Phi)]\Big\}\\
&\quad\times e^{-i\int_0^sds'[\bm{P}_{\perp}-\bm{\mathcal{A}}_{\perp}(\Phi+2s'P_{\tau})]^2}e^{2isP_{\tau}P_{\phi}}[\hat{\Pi}(\Phi)+m].
\end{split}
\end{equation}
Here, we have introduced the operator $\Phi$ of the light-cone time and the operator of the kinetic four-momentum $\Pi^{\mu}(\Phi)=P^{\mu}-\mathcal{A}^{\mu}(\Phi)$ in the plane wave, where $P^{\mu}$ is the canonical four-momentum. The light-cone components of $P^{\mu}=i\partial^{\mu}$ are given by $P_{\phi}=-i\partial_{\phi}=-(\tilde{n}P)=-(i\partial_t-i\partial_{x_n})/2$, $P_{\tau}=-i\partial_{\tau}=-(nP)=-(i\partial_t+i\partial_{x_n})$, and $\bm{P}_{\perp}=(P_{\perp,1},P_{\perp,2})=-i(\bm{a}_1\cdot\bm{\nabla},\bm{a}_2\cdot\bm{\nabla})$. If $\ket{p}$ is the eigenstate of the four-momentum operator $P^{\mu}$ with eigenvalue $p^{\mu}$, then, by defining $\braket{x|p}=\exp(-i(px))=\exp[-i(p_+\phi+p_-\tau-\bm{p}_{\perp}\cdot\bm{x}_{\perp})]$, it is $P_{\phi}\ket{p}=-p_+\ket{p}$, $P_{\tau}\ket{p}=-p_-\ket{p}$, and $\bm{P}_{\perp}\ket{p}=\bm{p}_{\perp}\ket{p}$. The operators $P_{\phi}$, $P_{\tau}$, and $\bm{P}_{\perp}$ are also the momenta conjugated to the light-cone coordinates in the sense that the commutator between the operator corresponding to each light-cone coordinate and the associated momentum operator is equal to the imaginary unit (all other possible commutators vanish): $[\Phi,P_{\phi}]=[T,P_{\tau}]=i$ and $[X_{\perp,j},P_{\perp,k}]=i\delta_{jk}$, which are equivalent to the manifestly covariant commutator $[X^{\mu},P^{\nu}]=-i\eta^{\mu\nu}$ ($T$ indicates the operator corresponding to the light-cone variable $\tau$).

\subsection{The one-loop mass operator}
The one-loop mass operator $M(x,x')$ in configuration space is defined via the equation
\begin{equation}
\label{M_xxp}
-iM(x,x')=-e^2D_{\mu\nu}(x-x')\gamma^{\mu}G(x,x')\gamma^{\nu},
\end{equation}
where
\begin{equation}
D^{\mu\nu}(x)=\int\frac{d^4q}{(2\pi)^4}\frac{\eta^{\mu\nu}}{q^2-\kappa^2+i0}e^{-i(qx)},
\end{equation}
is the photon propagator, with $\kappa^2$ being the square of the fictitious photon mass, which has been introduced to avoid infrared divergences \cite{Peskin_b_1995}.

We are interested in computing the one-loop mass operator $M(l,l')$ in momentum space, defined as
\begin{equation}
\label{M_llp}
M(l,l')=\int d^4xd^4x'\bar{E}(l,x)M(x,x')E(l',x').
\end{equation}
By substituting the expression of the photon propagator, we have [see Eq. (\ref{M_xxp})]
\begin{equation}
\begin{split}
M(l,l')&=-ie^2\int d^4xd^4x'\int\frac{d^4q}{(2\pi)^4}\frac{e^{-i(q(x-x'))}}{q^2-\kappa^2+i0}\bar{E}(l,x)\gamma^{\mu}G(x,x')\gamma_{\mu}E(l',x').
\end{split}
\end{equation}
Now, we exponentiate the denominator of the photon propagator in the usual way by introducing the Schwinger proper time $u$ and we use the operator form of the electron propagator according to Eq. (\ref{G}) such that the mass operator can be written as
\begin{equation}
\begin{split}
M(l,l')&=ie^2\int d^4x\int\frac{d^4q}{(2\pi)^4}\int_0^{\infty} du ds\,e^{iu(q^2-\kappa^2)-ism^2}\\
&\quad\times\bar{E}(l,x)\gamma^{\mu}\Big\{1+\frac{1}{2P_{\tau}}\hat{n}[\hat{\mathcal{A}}(\phi+2sP_{\tau}-2sq_-)-\hat{\mathcal{A}}(\phi)]\Big\}\\
&\quad\times e^{-i\int_0^sds'[\bm{P}_{\perp}+\bm{q}_{\perp}-\bm{\mathcal{A}}_{\perp}(\phi+2s'P_{\tau}-2s'q_-)]^2}e^{2is(P_{\tau}-q_-)(P_{\phi}-q_+)}[\hat{\Pi}(\Phi)+\hat{q}+m]\gamma_{\mu}E(l',x),
\end{split}
\end{equation}
where we have also exploited the operator relation $\exp(i(Xq))g(P)\exp(-i(Xq))=g(P+q)$, where $g(P)$ is a generic function of the four-momentum operator $P^{\mu}$, intended to be expanded in Taylor series.

As it is typical in problems in the presence of a background plane wave depending only on the light-cone time $\phi$, the action of the operators $\bm{P}_{\perp}$ and $P_{\tau}$ on the matrix $E(l,x)$ is trivial, as well as the integrals over the conjugated variables $\bm{x}_{\perp}$ and $\tau$, which provide delta functions enforcing the conservation laws $\bm{l}_{\perp}=\bm{l}'_{\perp}$ and $l_-=l'_-$:
\begin{equation}
\begin{split}
M(l,l')&=ie^2(2\pi)^3\delta^2(\bm{l}_{\perp}-\bm{l}'_{\perp})\delta(l_--l'_-)\int d\phi e^{-i(l'_+-l_+)\phi}\int\frac{d^4q}{(2\pi)^4}\int_0^{\infty} du ds\,e^{iu(q^2-\kappa^2)-ism^2}\\
&\quad\times e^{-i\int_0^sds'[\bm{l}_{\perp}+\bm{q}_{\perp}-\bm{\mathcal{A}}_{\perp}(\phi_{s'})]^2+2is(l_-+q_-)(q_++l'_+)-2i(l_-+q_-)\int_0^sds'\big[\frac{\bm{l}_{\perp}\cdot\bm{\mathcal{A}}_{\perp}(\phi_{s'})}{l_-}-\frac{\bm{\mathcal{A}}_{\perp}^2(\phi_{s'})}{2l_-}\big]}\\
&\quad\times\bigg[1-\frac{\hat{n}\hat{\mathcal{A}}(\phi)}{2l_-}\bigg]\Bigglb(\left\{2\hat{\pi}_{l'}(\phi_s)\bigg[1-\frac{\hat{n}\widehat{\Delta\mathcal{A}}(\phi_s)}{2(l_-+q_-)}\bigg]+\gamma^{\mu}\bigg[1-\frac{\hat{n}\widehat{\Delta\mathcal{A}}(\phi_s)}{2(l_-+q_-)}\bigg]\hat{q}\gamma_{\mu}\right\}\\
&\quad\times\bigg[1+\frac{\hat{n}\widehat{\Delta\mathcal{A}}(\phi_s)}{2l_-}\bigg]\bigg[1+\frac{\hat{n}\hat{\mathcal{A}}(\phi)}{2l_-}\bigg]+4\bigg[1+\frac{\hat{n}\widehat{\Delta\mathcal{A}}(\phi_s)}{2l_-}\bigg]\bigg[1+\frac{\hat{n}\hat{\mathcal{A}}(\phi)}{2l_-}\bigg](m-\hat{l'})\Biggrb),
\end{split}
\end{equation}
where $\phi_s=\phi-2s(l_-+q_-)$ and $\Delta \mathcal{A}^{\mu}(\phi_s)=\mathcal{A}^{\mu}(\phi_s)-\mathcal{A}^{\mu}(\phi)$ [$\widehat{\Delta\mathcal{A}}(\phi_s)=\gamma_{\mu}\Delta\mathcal{A}^{\mu}(\phi_s)$], and where we have also used Eq. (\ref{E_p_Comm}) and the relation
\begin{equation}
\Pi^{\lambda}(\phi)E(l,x)=\left[\pi^{\lambda}_l(\phi)+i\frac{\hat{n}\hat{\mathcal{A}}'(\phi)}{2l_-}n^{\lambda}\right]E(l,x).
\end{equation}
Here, the four-vector
\begin{equation}
\label{pi}
\pi^{\lambda}_l(\phi)=l^{\lambda}-\mathcal{A}^{\lambda}(\phi)+\frac{(l\mathcal{A}(\phi))}{l_-}n^{\lambda}-\frac{\mathcal{A}^2(\phi)}{2l_-}n^{\lambda}
\end{equation}
is the classical kinetic four-momentum of an electron in the plane wave $A^{\mu}(\phi)$, with $\lim_{\phi\to\pm\infty}\pi^{\lambda}_l(\phi)=l^{\lambda}$.

Now, we take the four-dimensional integral in $d^4q$ exactly as in Ref. \cite{Di_Piazza_2020_b} in the case of the one-loop vertex correction. In fact, the integral over $d^2\bm{q}_{\perp}$ is Gaussian and the integral over $dq_+$ provides either a delta function $\delta((u+s)q_-+sl_-)$ or its derivative, which can then be used to take the integral in $dq_-$. After straightforward manipulations of the preexponential matrix, we finally obtain
\begin{equation}
\label{M_llp_nr}
\begin{split}
M(l,l')&=(2\pi)^3\delta^2(\bm{l}_{\perp}-\bm{l}'_{\perp})\delta(l_--l'_-)\frac{\alpha}{2\pi}\int d\phi\, e^{-i(l'_+-l_+)\phi}\int_0^{\infty} \frac{du ds}{(u+s)^2}\\
&\quad\times e^{-i\kappa^2u-i\frac{s^2}{u+s}\big\{m^2+\int_0^1dy[\Delta\bm{\mathcal{A}}_{\perp}(\phi_{ys})]^2-\big[\int_0^1dy\Delta\bm{\mathcal{A}}_{\perp}(\phi_{ys})\big]^2\big\}+i\frac{us}{u+s}(l^{\prime\,2}-m^2)}\\
&\quad\times\Bigglb[\hat{\Pi}_{1,l'}(\phi_s)\bigg[1-\frac{s}{u}\frac{\hat{n}\widehat{\Delta\mathcal{A}}(\phi_s)}{2l_-}\bigg]\\
&\qquad+\frac{\hat{n}}{2(u+s)}\Bigglb(\frac{s(u+s)}{ul_-}\left\{\int_0^1dy[\Delta\bm{\mathcal{A}}_{\perp}(\phi_{ys})]^2-\bigg[\int_0^1dy\Delta\bm{\mathcal{A}}_{\perp}(\phi_{ys})\bigg]^2\right\}\\
&\qquad-\frac{s^2}{ul_-}\left[\Delta\bm{\mathcal{A}}_{\perp}(\phi_s)-\int_0^1dy\Delta\bm{\mathcal{A}}_{\perp}(\phi_{ys})\right]^2\Biggrb)\\
&\qquad+\frac{s}{u+s}\hat{\Pi}_{2,l'}(\phi_s)\bigg[1+\frac{2u+s}{u}\frac{\hat{n}\widehat{\Delta\mathcal{A}}(\phi_s)}{2l_-}\bigg]+2\bigg[1+\frac{\hat{n}\widehat{\Delta\mathcal{A}}(\phi_s)}{2l_-}\bigg](m-\hat{l'})\Biggrb],
\end{split}
\end{equation}
where we have defined the two four-vectors [see also Eq. (\ref{pi})]
\begin{align}
\Pi^{\lambda}_{1,l}(\phi_s)&=l^{\lambda}-\Delta\mathcal{A}^{\lambda}(\phi_s)+\frac{(l\Delta\mathcal{A}(\phi_s))}{l_-}n^{\lambda}-\frac{(\Delta\mathcal{A}(\phi_s))^2}{2l_-}n^{\lambda},\\
\Pi^{\lambda}_{2,l}(\phi_s)&=l^{\lambda}-\int_0^1dy\Delta\mathcal{A}^{\lambda}(\phi_{ys})+\frac{(l\int_0^1dy\Delta\mathcal{A}(\phi_{ys}))}{l_-}n^{\lambda}-\frac{\big(\int_0^1dy\Delta\mathcal{A}(\phi_{ys})\big)^2}{2l_-}n^{\lambda}.
\end{align}
Note that, by exploiting the delta function resulting from the integral in $dq_+$, we have that $\phi_s=\phi-2usl_-/(u+s)$ and $\phi_{ys}=\phi-2yusl_-/(u+s)$.

Apart from the infrared divergence cured via the introduction of the finite photon mass $\kappa$, it is known that the expression in Eq. (\ref{M_llp_nr}) of the mass operator $M(l,l')$ is ultraviolet divergent and requires renormalization \cite{Baier_1976_a,Ritus_1985}. The renormalization is carried out by adding and subtracting the vacuum expression of the mass operator and by renormalizing the latter exactly as in vacuum, because the difference between the mass operators in the plane wave and in vacuum is finite \cite{Baier_1976_a,Ritus_1970}. The resulting expression of the renormalized mass operator $M_R(l,l')$ is
\begin{equation}
\label{M_llp_r}
\begin{split}
M_R(l,l')&=(2\pi)^3\delta^2(\bm{l}_{\perp}-\bm{l}'_{\perp})\delta(l_--l'_-)\frac{\alpha}{2\pi}\int d\phi\, e^{-i(l'_+-l_+)\phi}\int_0^{\infty} \frac{du ds}{(u+s)^2}e^{-i\kappa^2u-i\frac{s^2}{u+s}m^2}\\
&\quad\times \Bigglb\{e^{-i\frac{s^2}{u+s}\big\{\int_0^1dy[\Delta\bm{\mathcal{A}}_{\perp}(\phi_{ys})]^2-\big[\int_0^1dy\Delta\bm{\mathcal{A}}_{\perp}(\phi_{ys})\big]^2\big\}+i\frac{us}{u+s}(l^{\prime\,2}-m^2)}\\
&\qquad\times\Bigglb[\hat{\Pi}_{1,l'}(\phi_s)\bigg[1-\frac{s}{u}\frac{\hat{n}\widehat{\Delta\mathcal{A}}(\phi_s)}{2l_-}\bigg]\\
&\quad\qquad+\frac{\hat{n}}{2(u+s)}\Bigglb(\frac{s(u+s)}{ul_-}\left\{\int_0^1dy[\Delta\bm{\mathcal{A}}_{\perp}(\phi_{ys})]^2-\bigg[\int_0^1dy\Delta\bm{\mathcal{A}}_{\perp}(\phi_{ys})\bigg]^2\right\}\\
&\quad\qquad-\frac{s^2}{ul_-}\left[\Delta\bm{\mathcal{A}}_{\perp}(\phi_s)-\int_0^1dy\Delta\bm{\mathcal{A}}_{\perp}(\phi_{ys})\right]^2\Biggrb)\\
&\quad\qquad+\frac{s}{u+s}\hat{\Pi}_{2,l'}(\phi_s)\bigg[1+\frac{2u+s}{u}\frac{\hat{n}\widehat{\Delta\mathcal{A}}(\phi_s)}{2l_-}\bigg]+2\bigg[1+\frac{\hat{n}\widehat{\Delta\mathcal{A}}(\phi_s)}{2l_-}\bigg](m-\hat{l'})\Biggrb]
\\
&\qquad\quad-\frac{u+2s}{u+s}m-\frac{u}{u+s}\left(1-2i\frac{u+2s}{u+s}m^2s\right)(m-\hat{l'})\Biggrb\}.
\end{split}
\end{equation}
This is our final expression of the renormalized mass operator in momentum space for off-shell four-momenta $l^{\mu}$ and $l^{\prime\,\mu}$. Note that the terms with three gamma matrices can be further reduced according to the identity
\begin{equation}
\label{abc}
\begin{split}
\hat{a}\hat{b}\hat{c}&=\frac{1}{4}\text{tr}(\gamma_{\mu}\hat{a}\hat{b}\hat{c})\gamma^{\mu}-\frac{1}{4}\text{tr}(\gamma^5\gamma_{\mu}\hat{a}\hat{b}\hat{c})\gamma^5\gamma^{\mu}\\
&=\hat{a}(bc)-\hat{b}(ac)+\hat{c}(ab)+i\varepsilon_{\mu\nu\lambda\rho}\gamma^5\gamma^{\mu}a^{\nu}b^{\lambda}c^{\rho},
\end{split}
\end{equation}
where $a^{\mu}$, $b^{\mu}$, and $c^{\mu}$ are three generic four-vectors and $\gamma^5=i\gamma^0\gamma^1\gamma^2\gamma^3$.

In Ref. \cite{Baier_1976_a} the operator form $M_R$ of the mass operator is presented in the off-shell case. After computing the quantity $\int d^4x\bar{E}(l,x)M_R E(l',x)$, we have ensured that it is in agreement with Eq. (\ref{M_llp_r}). The advantage of computing directly the mass operator in momentum space $M_R(l,l')$ is that matrix elements over on-shell states can be directly evaluated, as we will do in the next section, by sandwiching Eq. (\ref{M_llp_r}) between two free electron spinors. 

Finally, in the case of a constant-crossed field, i.e., for $\bm{A}(\phi)=-\bm{E}\phi$, it can be easily shown that our expression of $M_R(l,l')$ reduces to the corresponding expression in Ref. \cite{Ritus_1970}.

\section{The electron mass shift}
\label{EMS}
As we have anticipated in the previous section, the availability of the mass operator, allows for computing the electron mass shift in an arbitrary plane wave. To this end, we start from the Schwinger-Dyson equation
\begin{equation}
\label{SD}
\{\gamma^{\mu}[i\partial_{\mu}-\mathcal{A}_{\mu}(\phi)]-m\}\Psi(x)=\int d^4y\, M_R(x,y)\Psi(y),
\end{equation}
for the spinor $\Psi(x)$, where [see Eq. (\ref{M_llp})]
\begin{equation}
\label{M_R}
M_R(x,y)=\int \frac{d^4l}{(2\pi)^4}\frac{d^4l'}{(2\pi)^4}E(l,x)M_R(l,l')\bar{E}(l',y).
\end{equation}

We seek for a solution of Eq. (\ref{SD}) up to first order in $\alpha$ and for convenience we introduce the following notation. Concerning the four-momenta, we indicate as $\bm{p}_{\text{lc}}$ the three light-cone components $(\bm{p}_{\perp},p_-)$ and, concerning the coordinates, we indicate as $\bm{x}_{\text{lc}}$ the three light-cone coordinates $(\bm{x}_{\perp},\tau)$ [in the case of an on-shell electron, the fourth component of the four-momentum is $p_+=(m^2+\bm{p}_{\perp}^2)/2p_-$ and $p_-> 0$]. Thus, by exploiting the completeness of the Volkov states $U_{\sigma}(p,x)$ and $V_{\sigma}(p,x)$ \cite{Bergou_1980,Ritus_1985,Zakowicz_2005,Boca_2010,Di_Piazza_2018_d}, we expand the first-order solution $\Psi^{(1)}(x)$ as
\begin{equation}
\label{Psi_1}
\Psi^{(1)}(x)=\sum_{\rho}\int\frac{d^3\bm{q}_{\text{lc}}}{(2\pi)^3}\frac{1}{2q_-}[c^{(1)}_{\rho}(q,\phi)U_{\rho}(q,x)+d^{(1)}_{\rho}(q,\phi)V_{\rho}(q,x)],
\end{equation}
where the integral in $q_-$ goes from zero to infinity (note that we are employing here the completeness of the Volkov states on an hypersurface at fixed light-cone time $\phi$, which was studied in particular in Ref. \cite{Bergou_1980}). By imposing that at the lowest order, the solution reduces to the positive-energy state characterized by the on-shell four-momentum $p^{\mu}$ and by the spin quantum number $\sigma$, we have that the first-order in $\alpha$ coefficients $c^{(1)}_{\rho}(q,\phi)$ and $d^{(1)}_{\rho}(q,\phi)$ are given by
\begin{align}
\label{c_1}
c^{(1)}_{\rho}(q,\phi)&=2q_-(2\pi)^3\delta^3(\bm{q}_{\text{lc}}-\bm{p}_{\text{lc}})\delta_{\rho \sigma}+\delta c_{\rho}(q,\phi),\\
\label{d_1}
d^{(1)}_{\rho}(q,\phi)&=\delta d_{\rho}(q,\phi),
\end{align}
where the quantities $\delta c^{(1)}_{\rho}(q,\phi)$ and $\delta d^{(1)}_{\rho}(q,\phi)$ scale as $\alpha$.

By substituting the expansion in Eq. (\ref{Psi_1}) with the coefficients $c^{(1)}_{\rho}(q,\phi)$ and $d^{(1)}_{\rho}(q,\phi)$ in Eqs. (\ref{c_1}) and (\ref{d_1}) in the Schwinger-Dyson equation (\ref{SD}) and by keeping only up-to-linear terms in $\alpha$, we obtain
\begin{equation}
i\hat{n}\sum_{\rho}\int\frac{d^3\bm{q}_{\text{lc}}}{(2\pi)^3}\frac{1}{2q_-}\left[\frac{d\delta c_{\rho}(q,\phi_x)}{d\phi_x}U_{\rho}(q,x)+\frac{d\delta d_{\rho}(q,\phi_x)}{d\phi_x}V_{\rho}(q,x)\right]=\int d^4y\, M_R(x,y)U_{\sigma}(p,y).
\end{equation}
Now, we project this equation over the Volkov states and, by using the orthogonality relations 
\begin{align}
\int d^3\bm{x}_{\text{lc}}\bar{U}_{\rho}(q,x)\hat{n}U_{\rho'}(q',x)&=2q_-(2\pi)^3\delta^3(\bm{q}_{\text{lc}}-\bm{q}'_{\text{lc}})\delta_{\rho\rho'},\\
\int d^3\bm{x}_{\text{lc}}\bar{V}_{\rho}(q,x)\hat{n}U_{\rho'}(q',x)&=0,\\
\int d^3\bm{x}_{\text{lc}}\bar{V}_{\rho}(q,x)\hat{n}V_{\rho'}(q',x)&=2q_-(2\pi)^3\delta^3(\bm{q}_{\text{lc}}-\bm{q}'_{\text{lc}})\delta_{\rho\rho'},
\end{align}
we obtain the two equations
\begin{align}
i\frac{d\delta c_{\rho}(q,\phi_x)}{d\phi_x}&=\int d^3\bm{x}_{\text{lc}}\int d^4y\,\bar{U}_{\rho}(q,x) M_R(x,y)U_{\sigma}(p,y),\\
i\frac{d\delta d_{\rho}(q,\phi_x)}{d\phi_x}&=\int d^3\bm{x}_{\text{lc}}\int d^4y\,\bar{V}_{\rho}(q,x) M_R(x,y)U_{\sigma}(p,y).
\end{align}
The right-hand sides of these equations can be worked out by using the representation (\ref{M_R}) of the mass operator and by introducing the matrix $\mathcal{M}_R(l,l',\phi)$ such that
\begin{equation}
M_R(l,l')=(2\pi)^3\delta^3(\bm{l}_{\text{lc}}-\bm{l}'_{\text{lc}})\int d\phi\, e^{-i(l'_+-l_+)\phi}\mathcal{M}_R(l,l',\phi).
\end{equation}
By taking first the integrals in $d^3\bm{x}_{\text{lc}}$ and $d^4y$, by using the resulting delta functions to take the integrals in $d^3\bm{l}_{\text{lc}}$ and $d^4l'$, and then finally by taking the integral in $dl_+$ by using the fact that the matrix $\mathcal{M}_R(l,l',\phi)$ does not depend on $l_+$, we obtain the equations
\begin{align}
\label{dc_dphi}
i\frac{d\delta c_{\rho}(q,\phi_x)}{d\phi_x}&=2m(2\pi)^3\delta^3(\bm{q}_{\text{lc}}-\bm{p}_{\text{lc}})\mathcal{M}_{R,\rho\sigma}(p,\phi),\\
\label{dd_dphi}
i\frac{d\delta d_{\rho}(q,\phi_x)}{d\phi_x}&=0,
\end{align}
where $\mathcal{M}_{R,\sigma\sigma'}(p,\phi)=\bar{u}_{\sigma}(p)\mathcal{M}_R(p,p,\phi)u_{\sigma'}(p)/2m$ and where the second equality arises because the argument of the delta function $\delta(q_-+p_-)$ never vanishes. By using Eqs. (\ref{M_llp_r}) and (\ref{abc}) and by rescaling the integration variables as $m^2s\to s$ and $m^2u\to u$, it is easily shown that
\begin{align}
\label{M_sigma_sigma}
\begin{split}
&\frac{\mathcal{M}_{R,\pm\pm}(p,p,\phi)}{m}\\
&\quad=\frac{\alpha}{4\pi}\int_0^{\infty} \frac{du ds}{(u+s)^2}\Bigglb[2\frac{u+2s}{u+s}\Bigglb(e^{-i\frac{s^2}{u+s}\big\{1+\int_0^1dy[\Delta\bm{\xi}_{\perp}(\phi_{ys})]^2-\big[\int_0^1dy\Delta\bm{\xi}_{\perp}(\phi_{ys})\big]^2\big\}}-e^{-i\frac{s^2}{u+s}}\Biggrb)\\
&\qquad+e^{-i\frac{s^2}{u+s}\big\{1+\int_0^1dy[\Delta\bm{\xi}_{\perp}(\phi_{ys})]^2-\big[\int_0^1dy\Delta\bm{\xi}_{\perp}(\phi_{ys})\big]^2\big\}}\Bigg\{\frac{u+2s}{u+s}[\Delta\bm{\xi}_{\perp}(\phi_s)]^2+\frac{s}{u}\int_0^1dy[\Delta\bm{\xi}_{\perp}(\phi_{ys})]^2\\
&\qquad-2\frac{s}{u}\frac{s}{u+s}\left[\int_0^1dy\Delta\bm{\xi}_{\perp}(\phi_{ys})\right]^2-\frac{s}{u}\frac{2u-s}{u+s}\Delta\bm{\xi}_{\perp}(\phi_s)\cdot\int_0^1dy\Delta\bm{\xi}_{\perp}(\phi_{ys})\\
&\qquad+ i\frac{s}{u}\frac{1}{p_-}\left[\zeta_{\pm\pm,\mu}\Delta\tilde{\xi}^{\mu\nu}(\phi_s)\Pi_{1,p,\nu}(\phi_s)-\frac{2u+s}{u+s}\zeta_{\pm\pm,\mu}\Delta\tilde{\xi}^{\mu\nu}(\phi_s)\Pi_{2,p,\nu}(\phi_s)\right]\Bigg\}\Biggrb],
\end{split}\\
\label{M_sigma_msigma}
\begin{split}
&\frac{\mathcal{M}_{R,\pm\mp}}{m}=i\frac{\alpha}{4\pi}\int_0^{\infty} \frac{du ds}{(u+s)^2}e^{-i\frac{s^2}{u+s}\big\{1+\int_0^1dy[\Delta\bm{\xi}_{\perp}(\phi_{ys})]^2-\big[\int_0^1dy\Delta\bm{\xi}_{\perp}(\phi_{ys})\big]^2\big\}}\\
&\qquad\times\frac{s}{u}\frac{1}{p_-}\left[\zeta_{\pm\mp,\mu}\Delta\tilde{\xi}^{\mu\nu}(\phi_s)\Pi_{1,p,\nu}(\phi_s)-\frac{2u+s}{u+s}\zeta_{\pm\mp,\mu}\Delta\tilde{\xi}^{\mu\nu}(\phi_s)\Pi_{2,p,\nu}(\phi_s)\right],
\end{split}
\end{align}
where $\Delta\bm{\xi}_{\perp}(\phi)=\Delta\bm{\mathcal{A}}_{\perp}(\phi)/m$, $\zeta^{\mu}_{\sigma\sigma'}=-\bar{u}_{\sigma}(p)\gamma^5\gamma^{\mu}u_{\sigma'}(p)/2m$, and $\Delta\tilde{\xi}^{\mu\nu}(\phi_s)=[\tilde{\mathcal{A}}^{\mu\nu}(\phi_s)-\tilde{\mathcal{A}}^{\mu\nu}(\phi)]/m=(1/2)\varepsilon^{\mu\nu\lambda\rho}[\mathcal{A}_{\lambda\rho}(\phi_s)-\mathcal{A}_{\lambda\rho}(\phi)]/m$, with $\mathcal{A}^{\mu\nu}(\phi)=n^{\mu}\mathcal{A}^{\nu}(\phi)-n^{\nu}\mathcal{A}^{\mu}(\phi)$, and where $\phi_s=\phi-2usp_-/m^2(u+s)$ and  $\phi_{ys}=\phi-2yusp_-/m^2(u+s)$ (note that the fictitious photon mass can be set to zero).

Now, Eqs. (\ref{dc_dphi}) and (\ref{dd_dphi}) imply that we can set $\delta d_{\rho}(q,\phi_x)=0$. By imposing the initial condition $\delta c_{\rho}(q,0)=0$ and by indicating as $U_{\sigma}^{(1)}(p,x)$ the first-order solution $\Psi^{(1)}(x)$ corresponding to the above given initial conditions, we have that
\begin{equation}
\label{U_1}
U_{\sigma}^{(1)}(p,x)=\left[1-i\frac{m}{p_-}\int_0^{\phi}d\phi'\mathcal{M}_{R,\sigma\sigma}(p,\phi')\right]U_{\sigma}(p,x)-i\frac{m}{p_-}\int_0^{\phi}d\phi'\mathcal{M}_{R,-\sigma\sigma}(p,\phi')U_{-\sigma}(p,x).
\end{equation}
This approximated solution is valid under the condition that $(m/p_-)\left|\int_0^{\phi}d\phi'\mathcal{M}_{R,\sigma\sigma'}(p,\phi')\right|\ll 1$ for all $\sigma,\sigma'=\pm 1$. Also, we conclude that radiative corrections do not mix states with different momenta and states with positive and negative energies. However, states with the same momentum (and sign of the energy) but different spin quantum numbers are, in general, mixed. In the case of a constant crossed field plus a monochromatic plane wave of arbitrary polarization, the corresponding matrix $\mathcal{M}_{R,\sigma\sigma'}(p,\phi)$ can be diagonalized and one can determine the corresponding electron quasienergies \cite{Baier_1976_c}. 
 
The treatment significantly simplifies in the case of a linearly polarized plane wave and by choosing the spin quantization axis as the direction of the magnetic field of the plane wave in the (initial) rest frame of the electron. In this case, we set $\psi_2(\phi)=0$ and $\bm{A}(\phi)=\psi_1(\phi)\bm{a}_1=A_0\psi(\phi)\bm{a}_1$, with $A_0<0$ being related to the amplitude of the electric field of the plane wave. For example, in the case of a monochromatic plane wave with amplitude $E_0$ and angular frequency $\omega_0$, we would have $A_0=-E_0/\omega_0$ and, e.g., $\psi(\phi)=\cos(\omega_0\phi)$, whereas in the constant crossed field case, we would have $A_0=-E_0/\omega_0$ and $\psi(\phi)=\omega_0\phi$, such that the angular frequency $\omega_0$ cancels out, as it should in this case. In this way, the electromagnetic field tensor $F^{\mu\nu}(\phi)$ of the wave and its dual $\tilde{F}^{\mu\nu}(\phi)$ can be written as $F^{\mu\nu}(\phi)=A_0^{\mu\nu}\psi'(\phi)$ and $\tilde{F}^{\mu\nu}(\phi)=\tilde{A}_0^{\mu\nu}\psi'(\phi)$, with $A_0^{\mu\nu}=A_0(n^{\mu}a_1^{\nu}-n^{\nu}a_1^{\mu})$ and $\tilde{A}^{\mu\nu}_0=(1/2)\varepsilon^{\mu\nu\lambda\rho}A_{0,\lambda\rho}$. With the mentioned choice of the spin quantization axis, the constant spinor $u_{\sigma}(p)$ fulfills the eigenvalue equation $\gamma^5\hat{\zeta}u_{\sigma}(p)=\sigma u_{\sigma}(p)$, where $\zeta^{\mu}=-\sigma\bar{u}_{\sigma}(p)\gamma^5\gamma^{\mu}u_{\sigma}(p)/2m=-\tilde{A}_0^{\mu\nu}p_{\nu}/(p_-A_0)$ is the spin four-vector, with $(\zeta p)=0$ and $\zeta^2=-1$. In the rest frame of the electron and in the case of constant crossed field, the three-dimensional spin vector $\bm{\zeta}$ points in the same direction of the magnetic field. In this case, one can see that the matrix $\mathcal{M}_{R,\sigma\sigma'}(p,\phi)$ is diagonal, that the states in Eq. (\ref{U_1}) with different spin quantum numbers do not mix but the diagonal elements $\mathcal{M}_{R,\sigma\sigma}(p,\phi)$ explicitly depend on the spin quantum number, i.e., the mass operator is not proportional to the unity matrix. Also, from the validity condition $(m/p_-)\left|\int_0^{\phi}d\phi'\mathcal{M}_{R,\sigma\sigma'}(p,\phi')\right|\ll 1$, we can write $1-i(m/p_-)\int_0^{\phi}d\phi'\mathcal{M}_{R,\sigma\sigma}(p,\phi')\approx \exp[-i(m/p_-)\int_0^{\phi}d\phi'\mathcal{M}_{R,\sigma\sigma}(p,\phi')]$. Thus, the quantity $\mathcal{M}_{R,\sigma\sigma}(p,\phi)$ can be interpreted as the electron mass shift $\delta m_{\sigma}(p,\phi)$ [see also the phase in Eq. (\ref{E_p}) and observe that $(px)=p_+\phi+p_-\tau-\bm{p}_{\perp}\cdot\bm{x}_{\perp}$, with $p_+=(m^2+\bm{p}^2_{\perp})/2p_-$] and it is
\begin{equation}
\label{delta_m}
\begin{split}
\frac{\delta m_{\sigma}(p,\phi)}{m}&=\frac{\alpha}{4\pi}\int_0^{\infty} \frac{du ds}{(u+s)^2}\Bigglb[2\frac{u+2s}{u+s}\Bigglb(e^{-i\frac{s^2}{u+s}\big\{1+\int_0^1dy[\Delta\bm{\xi}_{\perp}(\phi_{ys})]^2-\big[\int_0^1dy\Delta\bm{\xi}_{\perp}(\phi_{ys})\big]^2\big\}}-e^{-i\frac{s^2}{u+s}}\Biggrb)\\
&\quad+e^{-i\frac{s^2}{u+s}\big\{1+\int_0^1dy[\Delta\bm{\xi}_{\perp}(\phi_{ys})]^2-\big[\int_0^1dy\Delta\bm{\xi}_{\perp}(\phi_{ys})\big]^2\big\}}\\
&\quad\times\Bigg\{\frac{u+2s}{u+s}[\Delta\bm{\xi}_{\perp}(\phi_s)]^2+\frac{s}{u}\int_0^1dy[\Delta\bm{\xi}_{\perp}(\phi_{ys})]^2-2\frac{s}{u}\frac{s}{u+s}\left[\int_0^1dy\Delta\bm{\xi}_{\perp}(\phi_{ys})\right]^2\\
&\quad-\frac{s}{u}\frac{2u-s}{u+s}\Delta\bm{\xi}_{\perp}(\phi_s)\cdot\int_0^1dy\Delta\bm{\xi}_{\perp}(\phi_{ys})+2i\sigma\frac{us^2}{(u+s)^2}\int_0^1dy\,\chi(\phi_{ys})\Bigg\}\Biggrb],
\end{split}
\end{equation}
where $\chi(\phi)=-p_-A_0\psi'(\phi)/mE_{cr}$ is the local quantum nonlinearity parameter. In the above derivation, we have used the fact that $\chi(\phi)=-(\zeta\tilde{\chi}(\phi))$, where $\tilde{\chi}^{\mu}(\phi)=\tilde{F}^{\mu\nu}(\phi)p_{\nu}/mE_{cr}$.

As a check of the above expression of the mass shift, we have ensured that in the LCFA, the mass shift $\delta m_{\sigma}(p,\phi)$ reduces to the corresponding local expression of the mass shift found in Refs. \cite{Ritus_1970,Baier_1971} in the case of a background constant crossed field. Indeed, by expanding the right-hand side of Eq. (\ref{delta_m}) for $\phi_s$ and $\phi_{ys}$ around $\phi$ up to the first order, one finds
\begin{equation}
\label{delta_m_LCFA}
\frac{\delta m^{(\text{LCFA})}_{\sigma}}{m}=\frac{\alpha}{2\pi}\int_0^{\infty} \frac{du\, dv}{(1+v)^3}e^{-iu\left[1+\frac{1}{3}\frac{\chi^2(\phi)}{v^2}u^2\right]}\left[\frac{5+7v+5v^2}{3}\frac{\chi^2(\phi)}{v^2}u+i\sigma\chi(\phi)\right],
\end{equation}
where, for notational simplicity, we have not indicated the functional dependence on $p^{\mu}$ and $\phi$. In this derivation the changes of variables $s=uv$ and the $u\to (1+v)u/v^2$ and the identity
\begin{equation}
\begin{split}
&\int_0^{\infty}\frac{du\,dv}{u(1+v)^2}\frac{1+2v}{1+v}\left\{e^{-iu\left[1+\frac{1}{3}\frac{\chi^2(\phi)}{v^2}u^2\right]}-e^{-iu}\right\}\\
&\qquad=-\frac{\chi^2(\phi)}{3}\int_0^{\infty}\frac{du\,dv}{(1+v)^2}e^{-iu\left[1+\frac{1}{3}\frac{\chi^2(\phi)}{v^2}u^2\right]}\frac{1+v-3v^2}{1+v}\frac{u}{v^2}
\end{split}
\end{equation}
have been used.

\subsection{On the anomalous magnetic moment of the electron}
\label{AMM}

In vacuum QED the derivation of the anomalous magnetic moment of the electron starts from the computation of the vertex correction, with the external photon providing the magnetic field interacting with the electron \cite{Peskin_b_1995}. In the present case of an electron interacting with a background plane wave, the magnetic field of the wave can already be exploited to derive the anomalous magnetic moment of the electron starting from the mass operator rather than from the more complicated vertex correction \cite{Di_Piazza_2020_b}. This has been already carried out in the case of a constant crossed field in Refs. \cite{Ritus_1970,Baier_1971} by computing the mass correction of the electron in the field and then by extracting the anomalous magnetic moment of the electron from the term in the mass correction proportional to the electron spin.

Within the LCFA the field-dependent anomalous gyromagnetic factor of the electron $\delta g^{(\text{LCFA})}=g^{(\text{LCFA})}-2$ is obtained by equating the real part of the spin-dependent mass shift with $-\delta\bm{\mu}^{(\text{LCFA})}\cdot\bm{B}_0(\phi)=\delta g^{(\text{LCFA})}\mu_B(\sigma/2)\bm{\zeta}\cdot\bm{B}_0(\phi)$, where $\delta\bm{\mu}^{(\text{LCFA})}$ is the anomalous magnetic moment of the electron within the LCFA, $\mu_B=|e|/2m$ is the Bohr magneton, $\bm{B}_0(\phi)$ is the magnetic field of the plane wave in the initial rest frame of the electron, and where we have taken into account that the spin and the magnetic moment of the electron are oppositely directed because the electron charge is negative. Now, we notice that $\zeta_{\mu}\tilde{F}^{\mu\nu}(\phi)p_{\nu}/m=-\bm{\zeta}\cdot\bm{B}_0(\phi)$ and, with the above choice of $\zeta_{\mu}$, that $\zeta_{\mu}\tilde{F}^{\mu\nu}(\phi)p_{\nu}/m=-E_{cr}\chi(\phi)=-m\chi(\phi)/2\mu_B$. By combining the above equations, we obtain
\begin{equation}
\label{delta_mu_LCFA}
\frac{\delta g^{(\text{LCFA})}}{2}=-\frac{\alpha}{\pi}\text{Im}\int_0^{\infty} \frac{du dv}{(1+v)^3} e^{-iu\left[1+\frac{1}{3}\frac{\chi^2(\phi)}{v^2}u^2\right]},
\end{equation}
which is the corresponding local expression of the anomalous gyromagnetic factor of the electron in a constant crossed field \cite{Ritus_1970,Baier_1971}. In particular, in the weak-field limit $\chi(\phi)\to 0$, we obtain Schwinger's result $\delta g_0/2=\alpha/2\pi$ (recall that the original prescription on the poles of the propagators, allows one to write $\int_0^{\infty}du\,e^{-iu}=-i$).

The derivation of the result in Eq. (\ref{delta_mu_LCFA}) within the LCFA indicates that in the general case of an arbitrary plane wave it is not even possible to introduce the concept of a local electron anomalous magnetic moment. In this case, the interaction between the electron and the plane wave is non-local [see Eq. (\ref{delta_m}) and, in particular, the last, spin-dependent term] and it is not simply proportional to the magnetic field of the plane wave in the rest frame of the electron evaluated at $\phi$.

\section{Conclusions}
In conclusion, we have obtained the expression of the mass operator in momentum space in the presence of an arbitrary plane wave and for an off-shell electron. We have used the obtained expression to compute the mass shift for an on-shell electron. By specializing to the case of a linearly polarized wave and by choosing the spin quantization axis along the magnetic field of the plane wave in the initial rest frame of the electron, a compact expression of the mass shift is derived, which reduces to the expression in a constant crossed field obtained previously in the literature. Finally, we have shown that a local expression of the anomalous magnetic moment of the electron can be extracted from the mass shift within the locally constant field approximation. However, beyond the locally constant field approximation, the interaction between the plane wave and the electron is non-local, which prevents a convenient description of the interaction of the magnetic field of the wave and the electron simply via a magnetic moment.

\acknowledgments{This publication is also supported by the Collaborative Research Centre 1225 funded by Deutsche Forschungsgemeinschaft (DFG, German Research Foundation) - Project-ID 273811115 - SFB 1225.}

%


\end{document}